\documentclass[12pt]{iopart}

\usepackage{amssymb}
\usepackage{bm}

\newcommand{\beq}{\begin{equation}}
\newcommand{\eeq}{\end{equation}}

\newcommand{\ben}{\begin{eqnarray}}
\newcommand{\een}{\end{eqnarray}}
\newcommand{\benn}{\begin{eqnarray*}}
\newcommand{\eenn}{\end{eqnarray*}}

\begin{document}

\title[Dirac projectors for Hamiltonian systems]{On the use of projectors for Hamiltonian systems and their relationship with Dirac brackets}

\author{C Chandre$^1$, L de Guillebon$^1$, A Back$^1$, E Tassi$^1$ and P J Morrison$^2$}
\address{$^1$Centre de Physique Th\'eorique, CNRS -- Aix-Marseille Universit\'e, Campus de Luminy, case 907, F-13288 Marseille cedex 09, France\\
$^2$Department of Physics and Institute for Fusion Studies, The University of Texas at Austin, Austin, TX 78712-1060, USA}
\ead{chandre@cpt.univ-mrs.fr}


\begin{abstract}
The role of projectors associated with Poisson brackets of constrained Hamiltonian systems is analyzed. Projectors act in two instances in a bracket: in the explicit dependence on the variables and in the computation of the functional derivatives. The role of these projectors is investigated by using Dirac's theory of constrained Hamiltonian systems.  Results are illustrated by  three examples taken from plasma physics: magnetohydrodynamics, the Vlasov-Maxwell system,   and the linear two-species Vlasov system with quasineutrality.   
\end{abstract}

\maketitle
\section{Introduction}

We consider an arbitrary Poisson bracket of a Poisson algebra of functionals of field variables ${\bm\chi}({\bf x})$  given by
\begin{equation}
\label{eqn:PBgene}
\{F,G\}=\int \!d^n x \, F_{\bm \chi} \cdot {\mathbb J}  ({\bm\chi}) \cdot G_{\bm\chi}\,,
\end{equation}
where ${\bf x}\in {\mathbb R}^n$,  ${\bm \chi}\colon{\mathbb R}^n\rightarrow{\mathbb R}^d$,  and $F_{\bm \chi} \cdot {\mathbb J} \cdot G_{\bm\chi}= F_{\chi^i}  \,{\mathbb J}^{ij}  \, G_{\chi^j}$ with repeated indices summed.  By Poisson algebra we mean a Lie algebra realization on functionals with an associative product of functionals that satisfies the Leibniz law.  Also, we assume that the resulting equations of motion given by $\dot{\bm\chi}=\{{\bm\chi},H\}$, for some Hamiltonian functional  $H[{\bm\chi}]$, possess a conservation law ${\cal Q}[{\bm\chi}]=0$, where ${\cal Q}$ is a functional of the field variables and their derivatives. Here we address the specific case where these conservation laws are obtained regardless of the choice of Hamiltonian $H$, so ${\cal Q}=0$ is an intrinsic property of the bracket of the  Poisson algebra. 

There are two ways to define such a constrained  Poisson algebra.  The usual way is to place a restriction on the set of field variables ${\bm\chi}$ in the Poisson algebra. However,  this definition raises the question of  how to appropriately compute the constrained functional derivatives $F_{\bm\chi}$. The second way is to define a Poisson algebra that does not include any constraint on the field variables and, consequently, there is no ambiguity in defining the functional derivatives --  conservation laws such as ${\cal Q}=0$ take the form of Casimir invariants.  

In this article we investigate the links between these two ways of defining constrained Hamiltonian structures, and we propose a way to lift  Poisson structures defined via the constrained field variables approach  to ones that have the constraints as Casimir invariants.   As can be expected, the difficulty resides in assuring the validity of the Jacobi identity. If we keep the same Poisson bracket but extended to the bigger algebra (the one without any constraint on the field variables),  then in general, the Poisson structure is only obtained when the constraint is satisfied, i.e., the Jacobi identity is satisfied conditionally when ${\cal Q}[{\bm\chi}]=0$. It turns out that one can remedy this limitation by modifying the bracket with the inclusion of suitable projectors that leave the functional derivatives unconstrained and guarantee the Jacobi identity unconditionally. We identify such projectors acting on the functional derivatives and on the explicit dependence of the bracket on ${\bm\chi}$. We discuss the various choices of projectors and highlight a particularly relevant one obtained from Dirac's theory of constrained Hamiltonian systems. 

In order to illustrate our purpose,  consider the relatively simple and common example, the vorticity equation of a compressible or incompressible fluid in ${\mathbb R}^3$.  The vorticity ${\bm\omega}=\nabla\times {\bf v}$, with ${\bf v}$  the velocity field, satisfies
\begin{equation}
\frac{\partial {\bm\omega}}{\partial t}=\nabla \times ({\bf v}\times {\bm\omega}).
\label{eqn:vort}
\end{equation}
In terms of a commonly used Poisson bracket (see, e.g., Ref.~\cite{zakh97}), 
\begin{equation}
\{F,G\}_0=\int d^3x \, {\bm\omega}\cdot (\nabla \times F_{\bm\omega})\times(\nabla\times G_{\bm\omega})\,,
\label{eqn:brackvort}
\end{equation}
Eq.~(\ref{eqn:vort}) has the from $\dot{F}=\{F,H\}_0$ with the Hamiltonian $H=\int d^3x \,v^2/2$. 
However, if one forgets about the constraint on the vector fields ${\bm\omega}$ or if one wants to lift the algebra of functionals of divergence-free ${\bm\omega}$ to the algebra of functionals of any vector field ${\bm\omega}$, then the bracket~(\ref{eqn:brackvort}) does not  satisfy the Jacobi identity.  This is easily seen by the following counter example:
\[
F_1=\frac1{2}\int d^3x \,{\bm \omega}\cdot\hat{\bf x}\, {y^2}\,,\quad
F_2=\frac1{2}\int d^3 x\, {\bm \omega}\cdot\hat{\bf y}\, {z^2}\,,\quad
F_3=\int d^3 x \,{\bm \omega}\cdot \hat{\bf z}\, x\,,
\]
which yields,
\begin{equation*}
\{F_1,\{F_2,F_3\}_0\}_0 + \circlearrowleft = - \int d^3 x \,{\bm \omega}\cdot\nabla(yz)\neq0\,.
\label{counter}
\end{equation*}
Evidently, the bracket~(\ref{eqn:brackvort}) satisfies the Jacobi identity only if $\nabla \cdot {\bm\omega}=0$.  We refer to such  Poisson brackets that only satisfy the Jacobi identity conditionally as {\em tainted brackets}. One of the questions we address in this article is how to correct a  tainted bracket so that it satisfies the Jacobi identity unconditionally. 
For this particular example, the correction is obtained by inserting a projection operator, following Ref.~\cite{chan11}, given by ${\cal P}_\perp=1-\nabla \Delta^{-1}\nabla\cdot$, so that it defines a new bracket 
$$
\{F,G\}=\int d^3 x \, ({\cal P} {\bm\omega}) \cdot (\nabla \times F_{\bm\omega})\times(\nabla\times G_{\bm\omega}).
$$
It is rather straightforward (see  Ref.~\cite{chan11}) to show that this bracket satisfies the Jacobi identity unconditionally. We notice that $\nabla \cdot {\bm\omega}$ is a Casimir invariant of the modified bracket, i.e.\ $\{\nabla\cdot {\bm\omega},G\}=0$ for any functional $G$. 

As mentioned above, projectors are not only useful to lift algebras so as to satisfy the Jacobi identity, they are also involved in the way functional derivatives are computed when the field variables are constrained. As an illustration, we consider the incompressible Euler equation for the velocity field ${\bf v}({\bf x},t)$, 
$$
\dot{\bf v}=-{\bf v}\cdot \nabla {\bf v} -\nabla P,
$$
where $P$ is determined by the constraint $\nabla\cdot{\bf v}=0$. This equation has a Hamiltonian structure~\cite{arno66,holm85,holm98,mars02} given by the Hamiltonian $H[{\bf v}]=\int d^3x \, v^2/2$ 
and the Poisson bracket
$$
\{F,G\}=\int d^3 x \,{\bf v}\cdot [F_{\bf v},G_{\bf v}]_L,
$$
where $F_{\bf v}$ are the functional derivatives of an observable $F$ with respect to the field variable ${\bf v}$  and the Lie bracket $[{\bf V},{\bf W}]_L$ is given by
$$
[{\bf V},{\bf W}]_L=({\bf W}\cdot \nabla){\bf V}-({\bf V}\cdot\nabla ){\bf W}.
$$
It should be noted that the incompressible Euler equation cannot be directly obtained from $\dot{F}=\{F,H\}$ using unconstrained functional derivatives $F_{\bf v}$ since $\nabla\cdot {\bf v}=0$ would not be conserved by the flow. One way of correcting the bracket is to use an orthogonal projector~\cite{holm85}.  For divergence-free fields, this orthogonal projector is again given by ${\cal P}_\perp=1-\nabla \Delta^{-1}\nabla \cdot$  (see also  Refs.~\cite{zakh97,chan11}). In other words, the constrained functional derivative $F_{\bf v}$ must be computed such that it satisfies $\nabla \cdot F_{\bf v}=0$. However,  the fundamental reason for this constraint on the functional derivative is unclear,  even though it yields the correct equation of motion. For a more general constraint ${\cal Q}[{\bm\chi}]=0$, is it still the orthogonal projector that  has to be used for the constrained functional derivatives? In addition, this projector is in general not unique. It therefore raises natural questions such as which is the most relevant projector and how is it obtained in a systematic way? 

In this article, we investigate  two possible placements of a projectors:  one is on the explicit dependence on the field variables, while the other is  on the computation of the functional derivatives.  We clarify the choice of the relevant projector by using Dirac's theory of constrained Hamiltonian systems. In order to prove the relevance of these projectors, we consider three examples taken from plasma physics.  The first one is magnetohydrodynamics (MHD), both compressible and incompressible, the second one is the Vlasov-Maxwell system,  and the third example involves semi-local constraints on linear Vlasov equations with two species. 

The goal of this paper is twofold: The first purpose is to show that using some relevant projector, the tainted brackets can be corrected such that the new brackets satisfy the Jacobi identity unconditionally. The second purpose is to connect these corrected brackets to the ones obtained from Dirac's theory of constrained Hamiltonian systems.

\section{Formulation of the general method}

\subsection{Projected functional derivatives}

At the outset  we assume that the bracket~(\ref{eqn:PBgene}) is a Poisson bracket on the algebra of functionals of ${\bm \chi}$, where ${\bm\chi}$ denotes a $d$-tuple of fields such that ${\cal Q}[ {\bm\chi}]({\bf x})=0$ and ${\cal Q}[{\bm\chi}]$ is function of ${\bm\chi}$ and its derivatives. These  fields will be referred to as ${\cal Q}$-free  fields. In this section, our aim is to get a corresponding Poisson bracket on the algebra of any functionals of ${\bm\chi}$, satisfying ${\cal Q}[{\bm\chi}]({\bf x})=0$ or not. The functional derivatives ${\bar F}_{\bm\chi}$ are defined in the following way: 
\begin{equation}
\delta F=\int d^n x \,{\bar F}_{\bm\chi}\cdot \delta{\bm\chi}, 
\label{eqn:delf}
\end{equation}
for all ${\cal Q}$-free $\delta {\bm\chi}$, which here means that $\hat{\cal Q}\cdot \delta {\bm\chi}=0$ where $\hat{\cal Q}$ is the Fr\'echet derivative of ${\cal Q}$  defined by
$$
{\cal Q}[{\bm\chi}+\delta {\bm\chi}]({\bf x})-{\cal Q}[{\bm\chi}]({\bf x})=\hat{\cal Q} \delta {\bm\chi}+O(\Vert \delta {\bm\chi}\Vert^2).
$$
This means that ${\bar F}_{\bm\chi}$ is not uniquely defined: it is arbitrary up to an element of $\mbox{Rg } \hat{\cal Q}^\dagger$, since $\int d^nx\, {\bar F}_{\bm\chi}\cdot\delta{\bm\chi}=\int d^nx \,({\bar F}_{\bm\chi}+\hat{\cal Q}^\dagger {\bf g})\cdot \delta {\bm\chi}$ where ${\bf g}$ is arbitrary. We define the constrained functional derivative $\bar{F}_{\bm\chi}$ from the unconstrained one $F_{\bm\chi}$ by  the following equation:
\begin{equation}
\label{eqn:1var}
\int d^n x \,\bar{F}_{\bm\chi}\cdot \delta {\bm\chi}=\int  d^n x \,F_{\bm\chi}\cdot \delta \bar{\bm\chi},
\end{equation}
where now $\delta \bar{\bm\chi}$ is the constrained (${\cal Q}$-free) variation  and $\delta {\bm\chi}$ the unconstrained one. 
For the unconstrained variation $\delta {\bm\chi}$, we use a linear operator ${\cal P}$ acting as $\delta \bar{\bm\chi}={\cal P}^\dagger \delta {\bm\chi}$ such that $\hat{\cal Q}{\cal P}^\dagger=0$. Moreover,  the range of this operator ${\cal P}^\dagger$ should be $\mbox{Ker } \hat{\cal Q}$ and, in addition, ${\cal P}^\dagger$ should act as the identity on $\mbox{ Ker } \hat{\cal Q}$. This is equivalent to requiring that ${\cal P}$ be a projector. Consequently, this  leads to a condition on the possible projectors ${\cal P}$ such that ${\bar F}_{\bm \chi}={\cal P}F_{\bm\chi}$, viz. 
\begin{equation}
\label{eqn:comm}
\mbox{Ker }{\cal P}= \mbox{Rg }\hat{\cal Q}^\dagger.
\end{equation} 
Note that given this condition, ${\cal Q}[{\bm\chi}]({\bf x})$ is a Casimir invariant that is  naturally preserved by the flow. 
Still this projector is not unique. In the literature (see, e.g., Ref.~\cite{holm85}), the functional derivative is chosen such that $\hat{\cal Q} F_{\bm\chi}=0$, so that the projector satisfies $\hat{\cal Q} {\cal P}=0$. This corresponds to the orthogonal projector
\begin{equation}
\label{eqn:orth}
{\cal P}_\perp=1-\hat{\cal Q}^\dagger (\hat{\cal Q}\hat{\cal Q}^\dagger)^{-1}\hat{\cal Q},
\end{equation}
provided  $\hat{\cal Q}\hat{\cal Q}^\dagger$ is invertible on $\mbox{Rg }\hat{\cal Q}$. However it is not clear if it is the best choice for the projection. Other solutions satisfy
\begin{eqnarray*}
&& {\cal P}_\perp {\cal P}={\cal P}_\perp, \\
&& {\cal P} {\cal P}_\perp ={\cal P},
\end{eqnarray*}   
which are needed in order to satisfy Eq.~(\ref{eqn:comm}).
Given a particular projector ${\cal P}$ the bracket~(\ref{eqn:PBgene}) becomes 
\begin{equation}
\label{eqn:PBp}
\{F,G\}_{\rm t}=\int d^n x \,({\cal P} F_{\bm\chi})\cdot {\mathbb J}({\bm\chi}) \cdot ({\cal P}  G_{\bm\chi}),
\end{equation}
where now the functional derivatives are the unconstrained ones. So we have released the constraint on the functional derivatives but, in general the Poisson bracket~(\ref{eqn:PBp}) does not satisfy the Jacobi identity for functionals of arbitrary ${\bm\chi}$, ones no longer restricted to ${\cal Q}$-free  fields, because ${\mathbb J}({\bm\chi})$ may give contributions that do not satisfy the Jacobi identify when ${\cal Q}[\bm\chi]\not= 0$. However, if the projector ${\cal P}$ does not depend on the field variables ${\bm\chi}$, as  is the case for the examples we deal with in this article, then 
a bracket that satisfies the Jacobi identity for all functionals of ${\bm\chi}$, satisfying ${\cal Q}[{\bm\chi}]=0$ or not, is given by
\begin{equation}
\label{eqn:PBproj}
\{F,G\}=\int d^n x \, ({\cal P}  F_{\bm\chi}) \cdot {\mathbb J}({\cal P} {\bm\chi})\cdot ({\cal P} G_{\bm\chi}).
\end{equation}
In order to verify the Jacobi identity, we perform the change of variables ${\bm \chi}_P={\cal P}{\bm\chi}$ and ${\bm \chi}_Q={\bm\chi}-{\cal P} {\bm\chi}$ so that bracket~(\ref{eqn:PBproj}) formally becomes bracket~(\ref{eqn:PBp}) with ${\bm\chi}_P$ instead of ${\bm\chi}$. Since ${\bm \chi}_P$ is by definition ${\cal Q}$-free, the Jacobi identity is satisfied. 
For the Poisson bracket~(\ref{eqn:PBproj}), we notice that ${\cal Q}[{\bm \chi}]({\bf x})$ is a Casimir invariant, and that the equations of motion for ${\bm\chi}_P$ are identical to the ones given by the Poisson bracket~(\ref{eqn:PBgene}) or (\ref{eqn:PBp}).

\subsection{Dirac brackets}

\subsubsection{Local constraints}
\label{local}

In order to identify the most appropriate projector, we use Dirac's theory of constrained Hamiltonian systems~\cite{Dira50}. We begin with the following good Poisson bracket: 
\begin{equation}
\label{eqn:PBg}
\{F,G\}=\int d^nx \,F_{\bm\chi}\cdot \mathbb{J}({\bm\chi})\cdot G_{\bm\chi}
\end{equation}
and then impose the local constraint $\Phi({\bf x}):= {\cal Q}[{\bm\chi}]({\bf x})=0$, where as before ${\cal Q}[{\bm\chi}]({\bf x})$ is a function of ${\bm \chi}({\bf x})$ and its derivatives. The Dirac procedure begins with the computation of the matrix of Poisson brackets between the local constraints, 
$$
C({\bf x},{\bf x}')\equiv \{\Phi({\bf x}),\Phi({\bf x}')\}=\hat{\cal Q}\mathbb{J}\hat{\cal Q}^\dagger \delta({\bf x}-{\bf x}').
$$ 
We set ${\cal A}:= \hat{\cal Q}\mathbb{J}\hat{\cal Q}^\dagger$ and we assume that this quantity is invertible. Then, the  Dirac correction to the bracket~(\ref{eqn:PBg}) is given by
$$
-\int \int d^n x \, d^nx' \, \{F,\Phi({\bf x})\}D({\bf x},{\bf x}')\{\Phi({\bf x}'),G\},
$$
where $D({\bf x},{\bf x}')={\cal A}^{-1}({\bm \chi}({\bf x}))\delta({\bf x}-{\bf x}')$.
Since $\{F,\Phi({\bf x})\}=-\hat{\cal Q}\mathbb{J}\cdot F_{\bm\chi}$, this contribution is equal to
$$
-\int d^n x \,F_{\bm\chi}\cdot \mathbb{J}\hat{\cal Q}^\dagger {\cal A}^{-1}\hat{\cal Q}\mathbb{J}\cdot  G_{\bm\chi}.
$$
Therefore,  the Dirac bracket is given by 
\begin{equation}
\label{eq:DB}
\{F,G\}_*=\int d^nx \,F_{\bm\chi}\cdot \mathbb{J}_*({\bm\chi})\cdot  G_{\bm\chi},
\end{equation}
where 
$$
\mathbb{J}_*=\mathbb{J}-\mathbb{J}\hat{\cal Q}^\dagger {\cal A}^{-1}\hat{\cal Q}\mathbb{J}.
$$
We notice that we only need to verify that ${\cal A}$ is invertible on the range of ${\cal Q}$ in order to define the Dirac bracket~(\ref{eq:DB}). It is straightforward to verify that $\mathbb{J}_*$ is antisymmetric because ${\cal A}$ is antisymmetric. 
We notice that $\hat{\cal Q}\mathbb{J}_*=0$ (and therefore $\mathbb{J}_*\hat{\cal Q}^\dagger=0$). As a consequence, the constraint $\Phi$ is a Casimir invariant. 
The Poisson brackets obtained by the Dirac procedure are Poisson brackets of the form~(\ref{eqn:PBp}) but untainted, i.e., they satisfy the Jacobi identity unconditionally even though they are not of the form~(\ref{eqn:PBproj}) in general. This can be seen by considering a projector ${\cal P}$ as discussed in the previous section. Under the assumption that $\mbox{Ker }{\cal P}=\mbox{Rg }\hat{\cal Q}^\dagger$, we deduce that ${\mathbb J}_*(1-{\cal P})=0$, and consequently:
$$
\mathbb{J}_*={\cal P}^\dagger \mathbb{J}_*{\cal P}. 
$$
With this equality, the Poisson bracket becomes 
$$
\{F,G\}_*=\int d^n x \,({\cal P} F_{\bm\chi})\cdot {\mathbb J}_*({\bm\chi}) \cdot ({\cal P} G_{\bm\chi}).
$$
The additional feature is that, a priori, the Poisson matrix $\mathbb{J}_*$ is a function of both ${\cal P}  {\bm \chi}$ and $(1-{\cal P}) {\bm\chi}$. However, it is straightforward to check that $(1-{\cal P}) {\bm\chi}$ is a Casimir invariant.

Dirac's procedure shows that among the possible projectors ${\cal P}$ satisfying Eq.~(\ref{eqn:comm}), one turns out to be most  convenient. The matrix ${\mathbb J}_*$ can be rewritten using the Dirac projector 
\begin{equation}
\label{eqn:defProjDir}
{\cal P}_*=1-\hat{\cal Q}^\dagger {\cal A}^{-1}\hat{\cal Q} {\mathbb J},
\end{equation}
as
$$
{\mathbb J}_*={\cal P}_*^\dagger {\mathbb J} {\cal P}_*, 
$$
so that the Dirac bracket becomes the same as the original one~(\ref{eqn:PBg}) with the exception that the functional derivatives  are projected using the Dirac projector, 
\begin{equation}
\label{eqn:compute_DB}
\{F,G\}_*=\int d^n x \,({\cal P}_*   F_{\bm\chi}) \cdot {\mathbb J}({\bm \chi})\cdot ({\cal P}_* G_{\bm\chi}),
\end{equation}
where we notice that the Poisson matrix is ${\mathbb J}$ and not ${\mathbb J}_*$. The main difference between the the orthogonal projector ${\cal P}_\perp$ and the Dirac projector ${\cal P}_*$ is that ${\cal P}_\perp$ is a purely geometric object since it only depends on the constraints, and ${\cal P}_*$ is a dynamical object since it involves the Poisson matrix.   

{\em Remark:} 
We observe that the matrix corresponding to the Dirac bracket has the following property:
$${\mathbb J}_*={\cal P}_*^\dagger {\mathbb J} {\cal P}_*= {\mathbb J} {\cal P}_*= {\cal P}_*^\dagger {\mathbb J}, $$
i.e., the Dirac bracket can be rewritten from Eq.~(\ref{eqn:compute_DB}) using only one Dirac projector, e.g., 
\begin{eqnarray*}
\{F,G\}_*= \int d^n x   \, F_{\bm\chi}  \cdot {\mathbb J}({\bm \chi}) \cdot {\cal P}_* G_{\bm\chi}.
\end{eqnarray*}
As a result, the computation of the Dirac bracket is made easier.

\subsubsection{Semi-local constraints}

The calculation of Sec.~\ref{local}  can be generalized to  allow semi-local constraints in phase space.  To this end we split the set of coordinates into two pieces, i.e., ${\bf x}=({\bf x}_1,{\bf x}_2)$ where ${\bf x}_1\in {\mathbb R}^{n-m}$ and ${\bf x}_2\in {\mathbb R}^{m}$. The semi-local constraints are given by
$$
\Phi({\bf x}_1)=\bar{\cal Q}[{\bm \chi}]({\bf x}_1)=\int d^m x_2\, {\cal Q}[{\bm \chi}]({\bf x}),
$$
where ${\cal Q}[{\bm \chi}]({\bf x})$ is a function of ${\bm \chi}({\bf x})$ and its derivatives. 
The linear operator $\hat{\bar{\cal Q}}$ is defined by the linear operator associated with the function ${\cal Q}$ by
$$
{\hat{\bar {\cal Q}}}=\int d^m x_2 \,\hat{{\cal Q}}.
$$
Since ${\hat{\bar {\cal Q}}}$ acting on a function of ${\bf x}$ is only a function of ${\bf x}_1$, the linear operator $\hat{\bar{\cal Q}}^\dagger$ is defined by  the equation
$$
\int d^{n-m }x_1 \,{\hat{\bar {\cal Q}}} {\bm\chi} \cdot {\bf w}({\bf x}_1)=\int d^n x \,{\bm\chi}({\bf x}) \cdot 
\hat{\bar{\cal Q}}^\dagger {\bf w}. 
$$
Consequently, $\hat{\bar {\cal Q}}^\dagger$ is a linear operator acting on functions of ${\bf x}_1$ as ${\hat{{\cal Q}}}^\dagger$, i.e., ${\hat{\bar {\cal Q}}}^\dagger w({\bf x}_1)={\hat{ {\cal Q}}}^\dagger w({\bf x}_1)$.
In a manner similar to that of  Sec.~\ref{local}, the computation of the Dirac bracket shows that the  operator  
$$
{\cal A}=\hat{\bar {\cal Q}} {\mathbb J} \hat{\bar {\cal Q}}^\dagger ,
$$
must be invertible. 
More explicitly, the linear operator ${\cal A}$ acts on functions of ${\bf x}_1$ as 
$$
{\cal A} w({\bf x}_1)=\int d^m x_2 {\hat{\cal Q}}{\mathbb J} {\hat{\cal Q}}^\dagger w({\bf x}_1). 
$$
The expression of the Dirac projector is given by
$$
{\cal P}_*=1- {\hat{\bar {\cal Q}}}^\dagger {\cal A}^{-1} {\hat{\bar {\cal Q}}} {\mathbb J},
$$
in a very similar way as the case of the local constraints. We notice that the linear operator ${\cal A}$ only needs to be invertible on $\mbox{ Rg } {\hat{\bar {\cal Q}}}$. Another important projector is the orthogonal projector given by
$$
{\cal P}_\perp= 1-{\hat{\bar {\cal Q}}}^\dagger ({\hat{\bar {\cal Q}}}{\hat{\bar {\cal Q}}}^\dagger)^{-1}{\hat{\bar {\cal Q}}}. 
$$
As in the case of local constraints, these two projectors satisfy ${\mathbb J}_*={\cal P}^\dagger {\mathbb J}_* {\cal P}$, along with  the two properties ${\cal P}_\perp {\cal P}_*={\cal P}_\perp$ and ${\cal P}_*{\cal P}_\perp={\cal P}_*$. In addition, the Dirac projector satisfies ${\mathbb J}_*={\cal P}_*^\dagger {\mathbb J}{\cal P}_*={\cal P}_*^\dagger {\mathbb J}={\mathbb J}{\cal P}_*$. 

\section{Example 1: magnetohydrodynamics}

\subsection{Compressible magnetohydrodynamics}
\label{compMHD}

A particularly interesting example is afforded by the Hamiltonian structure of magnetohydrodynamics. The equations for the velocity field ${\bf v}({\bf x},t)$, the density $\rho({\bf x},t)$, the magnetic field ${\bf B}({\bf x},t)$,  and the entropy $s({\bf x},t)$ are given by
\begin{eqnarray*}
&& \dot{\rho}=-\nabla\cdot (\rho{\bf v}),\\
&& \dot{\bf v}=-{\bf v}\cdot\nabla{\bf v}-\rho^{-1}\nabla (\rho^2 U_\rho) +\rho^{-1} (\nabla \times{\bf B})\times {\bf B},\\
&& \dot{\bf B}=\nabla\times({\bf v}\times{\bf B}),\\
&& \dot{s}=-{\bf v}\cdot\nabla s,
\end{eqnarray*}
where $U$ is the internal energy and $U_\rho$ here denotes the partial derivative of $U$ with respect to $\rho$.
The dynamical variables are $\rho({\bf x})$, ${\bf v}({\bf x})$, ${\bf B}({\bf x})$ and $s({\bf x})$ where ${\bf x}$ belongs to ${\mathbb R}^3$. The observables of the system are functionals of these vector fields, denoted generically by $F(\rho,{\bf v},{\bf B},s)$.
In these coordinates, this system has the following Hamiltonian
\begin{equation*}
\label{eqn:H}
H(\rho,{\bf v},{\bf B},s)=\int d^3x\left( \frac{1}{2}\rho{\bf v}^2 +\rho U(\rho,s)+\frac{{\bf B}^2}{2}\right).
\end{equation*}
There are two slightly different Poisson brackets that have been proposed in Refs.~\cite{morr80a,morr82e,morr82}.
A first one was given in Ref.~\cite{morr80a}, 
\begin{eqnarray}
\{F,G\}&=&-\int d^3x \,\left[ F_\rho \nabla\cdot G_{\bf v} +F_{\bf v}\cdot\nabla G_\rho -\rho^{-1}(\nabla\times {\bf v})\cdot \left( F_{\bf v}\times G_{\bf v}\right)\right.\nonumber \\
&&\qquad \qquad \left.+\rho^{-1}\nabla s \cdot \left( F_s G_{\bf v}-F_{\bf v} G_s\right)\right] +\{F,G\}_B, \label{eqn:PB1}
\end{eqnarray}
where the magnetic part $\{F,G\}_B$ of the Poisson bracket is chosen as $\{F,G\}_B=\{F,G\}_{B,{\rm t}}$
\begin{equation}
\{F,G\}_{B,{\rm t}}=-\int d^3x \, \rho^{-1}\left( F_{\bf v}\cdot{\bf B}\times( \nabla\times G_{\bf B})-G_{\bf v}\cdot{\bf B}\times( \nabla\times F_{\bf B})\right). \label{eq:PBmpt}
\end{equation}
It was pointed out in  Ref.~\cite{morr82} that this bracket satisfies the Jacobi identity only when $\nabla\cdot {\bf B}=0$, and also that  $\nabla\cdot {\bf B}$ commutes with any other functionals, i.e., $\{F,\nabla\cdot {\bf B}\}=0$ for all $F$ (it is a Casimir-like property, even though we cannot call it a Casimir invariant since the Jacobi identity is only satisfied when $\nabla\cdot {\bf B}=0$). As was the case for the vorticity equation~(\ref{eqn:vort}), the functional derivatives with respect to ${\bf B}$ must be divergence-free for coherence. However,  we notice that here, since only $\nabla\times F_{\bf B}$ are involved in the expression of the magnetic part~(\ref{eqn:PB1}) of the Poisson bracket, it does not make any difference whether $F_{\bf B}$ is divergence-free or not. 

In order to extend the definition of the Poisson bracket to functionals of any ${\bf B}$, ones not necessarily divergence-free, a second Poisson bracket was proposed in Refs.~\cite{morr82e,morr82}. There the magnetic part of the Poisson bracket~(\ref{eqn:PB1}) was replaced by
\begin{eqnarray*}
\{F,G\}_{B,1}&=& -\int d^3x \,\left[\left( \rho^{-1}F_{\bf v} \cdot [\nabla G_{\bf B}] - \rho^{-1}G_{\bf v} \cdot [\nabla F_{\bf B}]\right) \cdot {\bf B}\right. \\
&& \qquad \qquad \left.+ {\bf B}\cdot \left( [\nabla \left(\rho^{-1}F_{\bf v}\right)]\cdot G_{\bf B}- [\nabla \left(\rho^{-1}G_{\bf v}\right)] \cdot F_{\bf B}\right)\right]. \label{eqn:PB}
\end{eqnarray*}
Here the notation ${\bf a}\cdot [M] \cdot{\bf b}$ is a scalar explicitly given by $\sum_{ij}a_i M_{ij} b_j$ for any vectors ${\bf a}$ and ${\bf b}$ and any matrix $[M]$. It was shown that this bracket satisfies the Jacobi identity for all functionals of $(\rho,{\bf v},s,{\bf B})$ regardless of the condition $\nabla\cdot{\bf B}=0$.
The magnetic part of this Poisson bracket is rewritten as 
\begin{eqnarray}
\{F,G\}_{B,1}&=&-\int d^3x\,\rho^{-1}\left[ F_{\bf v}\cdot{\bf B}\times( \nabla\times G_{\bf B})-G_{\bf v}\cdot{\bf B}\times( \nabla\times F_{\bf B})\right] \nonumber \\
&& +\int d^3x \,\rho^{-1}\nabla \cdot {\bf B} \left( F_{\bf v}\cdot G_{\bf B}-F_{\bf B}\cdot G_{\bf v}\right).\label{eq:MHDPB2}
\end{eqnarray}
The first line of the above bracket corresponds to the Poisson bracket introduced in Ref.~\cite{morr80a} [see Eq.~(\ref{eq:PBmpt})]. With the additional terms (proportional to $\nabla\cdot {\bf B}$) the Jacobi identity is unconditionally satisfied for any functionals of $(\rho,{\bf v},s,{\bf B})$. However, a property of the bracket~(\ref{eqn:PB1}) with the magnetic part~(\ref{eq:PBmpt}) has been lost, $\nabla\cdot {\bf B}$ does not Poisson-commute with any functional, so it is not a Casimir invariant. 

In order to have both the Jacobi identity unconditionally satisfied and $\nabla\cdot {\bf B}$ a Casimir invariant, we apply the prescription~(\ref{eqn:PBproj}) on the magnetic part~(\ref{eq:PBmpt}).   At every instance in the Poisson bracket where ${\bf B}$ is explicitly mentioned, we replace ${\bf B}$ with
$\bar{\bf B}={\bf B}-\nabla\Delta^{-1}\nabla\cdot {\bf B}$. The magnetic part becomes
$$
\{F,G\}_B=-\int d^3x\,\rho^{-1}\left( F_{\bf v}\cdot(\bar{\bf B}\times( \nabla\times G_{\bf B})-G_{\bf v}\cdot(\bar{\bf B}\times( \nabla\times F_{\bf B}))\right),
$$
and it is rewritten as
\begin{eqnarray}
&& \{F,G\}_B =-\int d^3x\,\rho^{-1}\left( F_{\bf v}\cdot({\bf B}\times( \nabla\times G_{\bf B})-G_{\bf v}\cdot({\bf B}\times( \nabla\times F_{\bf B}))\right)\nonumber  \\
&& \quad +\int d^3x \,\nabla \cdot {\bf B} \, \Delta^{-1}\nabla\cdot\left(\rho^{-1}F_{\bf v}\times(\nabla\times G_{\bf B})-\rho^{-1}G_{\bf v}\times(\nabla\times F_{\bf B})\right).\label{eq:MHDPB3}
\end{eqnarray}
Here we notice that the correction term still contains terms proportional to $\nabla \cdot {\bf B}$ but is different from the one in Eq.~(\ref{eq:MHDPB2}). The main difference is that $\nabla\cdot {\bf B}$ is not a Casimir invariant for the Poisson bracket~(\ref{eq:MHDPB2}) whereas it is one for the Poisson bracket~(\ref{eq:MHDPB3}) since it only involves terms like $\nabla\times G_{\bf B}$.

\subsection{Incompressible magnetohydrodynamics}
\label{sec:MHDincomp}

For incompressible MHD we begin with  the equations for  compressible magnetohydrodynamics from Sec.~\ref{compMHD} and apply constraints. The Poisson bracket given by Eqs.~(\ref{eqn:PB1})-(\ref{eq:MHDPB3}) is of the form~(\ref{eqn:PBg}) with 
$$
{\mathbb J}=\left( \begin{array}{cccc}
0 & -\nabla \cdot & 0 & 0\\
-\nabla & -\rho^{-1}(\nabla \times {\bf v})\times & -\rho^{-1}\bar{\bf B} \times (\nabla \times) & \rho^{-1} \nabla s  \\
0 & -\nabla\times (\rho^{-1} \bar{\bf B}\times) & 0 & 0\\
0 & -\rho^{-1}\nabla s & 0 & 0
\end{array}\right).
$$
We impose the following local constraints on the field variables ${\bm\chi}=(\rho,{\bf v}, {\bf B},s)$, 
$$
{\cal Q}[{\bm\chi}]({\bf x})=(\rho,\nabla\cdot {\bf v}). 
$$
The reduction to incompressible MHD using Dirac's theory has already been done in Ref.~\cite{chan11} and the reduction to incompressible Euler equation in Refs.~\cite{nguy99,nguy01}. Here we propose a more compact way to present this reduction using the operators introduced in the previous sections. The expressions of the intermediate operators are
\begin{eqnarray*}
&& \hat{\cal Q}=\left(\begin{array}{cccc} 1 & 0 & 0 & 0\\
0 & \nabla \cdot & 0 & 0
\end{array}\right),\\
&& \hat{\cal Q}^\dagger = \left(
\begin{array}{cc} 1 & 0\\
0 & -\nabla \\
0 & 0\\
0 & 0
\end{array}\right),\\
&& {\cal A}=\left( \begin{array}{cc} 0 & \Delta \\ -\Delta & \nabla \cdot (\rho^{-1}(\nabla\times {\bf v})\times \nabla) \end{array}\right),\\
&& {\cal A}^{-1}=\left( \begin{array}{cc} \Delta^{-1}\nabla \cdot (\rho^{-1}(\nabla\times {\bf v})\times \nabla) & -\Delta^{-1} \\ \Delta^{-1} & 0 \end{array}\right).
\end{eqnarray*}
The orthogonal projector is given by Eq.~(\ref{eqn:orth}) and its expression is 
$$
{\cal P}_\perp F_{\bm \chi}=(0,\bar{F}_{\bf v}, F_{\bf B}, F_s), 
$$
where $\bar{F}_{\bf v}=F_{\bf v}-\nabla \Delta^{-1}\nabla\cdot F_{\bf v}$.
The Dirac projector, computed from the Poisson bracket~(\ref{eqn:PB1}) where ${\bf B}$ has been replaced by $\bar{\bf B}={\bf B}-\nabla \Delta^{-1}\nabla \cdot {\bf B}$, is given by
$$
{\cal P}_*F_{\bm\chi}=(F_*,\bar{F}_{\bf v}, F_{\bf B}, F_s),
$$
where 
$$
F_*=\Delta^{-1}\nabla\cdot\left( \rho^{-1}\left( (\nabla\times{\bf v})\times \bar{F}_{\bf v}-\bar{\bf B}\times (\nabla \times F_{\bf B})- F_s \nabla s \right) \right).
$$
We notice that the two projectors differ in the first component. Even though the two projectors ${\cal P}_\perp$ and ${\cal P}_*$ are different, both of these projectors satisfy the equation ${\mathbb J}_*={\cal P}^\dagger {\mathbb J}{\cal P}$, which is always the case for the Dirac projector but not true in general for the orthogonal projector. Actually any projector ${\cal P}F_{\bm\chi}=(F_*(\bar{F}_{\bf v},F_{\bf B},F_s), \bar{F}_{\bf v},F_{\bf B},F_s)$ satisfies ${\mathbb J}_*={\cal P}^\dagger {\mathbb J}{\cal P}$ for any function $F_*$. The first component is thus irrelevant, and consequently the orthogonal projector is the simplest projector to be used for constrained functional derivatives. 
From this projector, we compute the Dirac bracket from Eq.~(\ref{eqn:compute_DB}), and it gives the same bracket as that produced  in Ref.~\cite{chan11}:
\begin{eqnarray*}
\{F,G\}_*&=& \int d^3 x \,\rho^{-1}\left( (\nabla\times {\bf v})\cdot ( \bar{F}_{\bf v}\times \bar{G}_{\bf v}) -\nabla s \cdot ( F_s \bar{G}_{\bf v}-\bar{F}_{\bf v} G_s)\right.\\
&& \qquad \qquad \left. + \bar{\bf B} \cdot ( \bar{F}_{\bf v}\times( \nabla\times G_{\bf B})+ ( \nabla\times F_{\bf B})\times \bar{G}_{\bf v})\right),
\end{eqnarray*}
where $\bar{F}_{\bf v}=F_{\bf v}-\nabla\Delta^{-1}\nabla\cdot F_{\bf v}$.

\section{Example 2: Vlasov-Maxwell equations}

\subsection{Vlasov-Maxwell modified bracket as a Dirac bracket}

As a second example, we consider the Vlasov-Maxwell equations for the distribution of charged particles in phase space $f({\bf x},{\bf v},t)$ and the electromagnetic fields ${\bf E}({\bf x},t)$ and ${\bf B}({\bf x},t)$ given by
\begin{eqnarray*}
&& \dot{f}=-{\bf v}\cdot \nabla f -({\bf E}+{\bf v}\times {\bf B})\cdot \partial_{\bf v} f,\\
&& {\dot {\bf E}}=\nabla \times {\bf B}-{\bf J},\\
&& \dot{\bf B}=-\nabla \times {\bf E},
\end{eqnarray*}
where ${\bf J}=\int d^3v \,{\bf v} f$. The Hamiltonian of this system is given by
$$
H=\int d^6z \,f \frac{{\bf v}^2}{2} +\int d^3x\,\frac{{\bf E}^2+{\bf B}^2}{2},
$$
where we denote ${\bf z}=({\bf x},{\bf v})$. The Poisson bracket between two functionals of $f({\bf x},{\bf v})$, ${\bf E}({\bf x})$ and ${\bf B}({\bf x})$ is given by
\begin{eqnarray}
\{F,G\}_{\rm t}&=&\int d^6 z \, f \left( [F_f,G_f]_{\rm c}+[F_f,G_f]_{\rm B}+G_{\bf E}\cdot \partial_{\bf v}F_f-F_{\bf E}\cdot \partial_{\bf v}G_f\right)\nonumber \\
&& +\int d^3x \,\left( F_{\bf E} \cdot\nabla \times G_{\bf B}-\nabla \times F_{\bf B}\cdot G_{\bf E}\right),
\label{eq:PBMV}
\end{eqnarray}
where the two brackets $[\cdot ,\cdot]_{\rm c}$ and $[\cdot ,\cdot ]_{\rm B}$ are defined by
\begin{eqnarray}
&& [f,g]_{\rm c}=\nabla f\cdot \partial_{\bf v}g-\partial_{\bf v}f\cdot \nabla g,\label{eqn:bc}\\
&& [f,g]_{\rm B}={\bf B}\cdot(\partial_{\bf v}f\times\partial_{\bf v}g). \label{eqn:bbp}
\end{eqnarray}
The Poisson bracket~(\ref{eq:PBMV}) was given in Ref.~\cite{mars82} based on an earlier work of Ref.~\cite{morr80b} (see also Ref.~\cite{bial84}).  It was pointed out in  Ref.~\cite{morr82} that the Poisson bracket~(\ref{eq:PBMV}) only satisfies the Jacobi identity when $\nabla \cdot {\bf B}=0$, which is to say that it does not satisfy the Jacobi identity for arbitrary functionals of $(f,{\bf E},{\bf B})$. This problem is actually already present  in the Lagrangian description (for the dynamics of charged particles) since $[\cdot,\cdot]_{\rm c}+[\cdot,\cdot]_{\rm B}$  only satisfies the Jacobi identity for functions ${\bf B}$ such that $\nabla\cdot {\bf B}=0$, whereas, individually, $[\cdot,\cdot]_{\rm c}$ and $[\cdot,\cdot]_{\rm B}$ satisfy the Jacobi identity for an arbitrary function ${\bf B}$.

In order to remedy this problem, we modify the bracket $[\cdot,\cdot]_{\rm B}$ to take the form of (\ref{eqn:PBproj}), 
$$
[f,g]_{{\rm B}_P}=({\bf B}-\nabla \Delta^{-1}\nabla\cdot {\bf B})\cdot(\partial_{\bf v}f\times\partial_{\bf v}g).
$$
With this modified gyrobracket, we readily check that $[\cdot,\cdot]_{\rm c}+[\cdot,\cdot]_{{\rm B}_P}$ satisfies the Jacobi identity. Next, we consider the modified Poisson bracket~(\ref{eq:PBMV}) obtained by replacing $[\cdot,\cdot]_{\rm B}$ by $[\cdot ,\cdot ]_{{\rm B}_P}$, i.e., we consider the Poisson bracket
\begin{eqnarray}
\{F,G\}_{\rm VM}&=&\int d^6 z \,f \left( [F_f,G_f]_{\rm c}+[F_f,G_f]_{{\rm B}_P}+G_{\bf E}\cdot \partial_{\bf v}F_f-F_{\bf E}\cdot \partial_{\bf v}G_f\right)\nonumber \\
&& +\int d^3x \,\left( F_{\bf E} \cdot\nabla \times G_{\bf B}-\nabla \times F_{\bf B}\cdot G_{\bf E}\right),
\label{eq:PBMVproj}
\end{eqnarray}
which satisfies the Jacobi identity unconditionally. This follows from the change of variable
${\bf B}_P={\bf B}-\nabla \Delta^{-1}\nabla\cdot {\bf B}$ and ${\bf B}_Q=\nabla \Delta^{-1}\nabla\cdot {\bf B}$
where it should be noted that 
$$
\nabla\times G_{\bf B}=\nabla \times G_{{\bf B}_P}, 
$$
since the operator ${\cal P}=1-\nabla\Delta^{-1}\nabla \cdot $ satisfies ${\cal P}\nabla\times =\nabla\times$.

Here it should be noticed that $\nabla\cdot {\bf B}$ is a Casimir invariant for the Poisson bracket~(\ref{eq:PBMVproj}). The untainted form of the Vlasov-Maxwell bracket~(\ref{eq:PBMVproj}) gives the Hamiltonian structure of the Vlasov-Maxwell equations in terms of physical fields without introducing the vector potential, i.e., without the restriction of $\nabla\cdot{\bf B}=0$.

In order to realize the link between brackets defined using projectors and Dirac brackets, we show below that the Poisson bracket~(\ref{eq:PBMVproj}) is a Dirac bracket of some parent bracket obtained using two constraints which, by definition, are Casimir invariants of the bracket~(\ref{eq:PBMVproj})
\begin{eqnarray*}
&& {\cal Q}[f,{\bf E},{\bf B}]({\bf x})=(\nabla\cdot {\bf E}-\rho, \nabla\cdot {\bf B}),
\end{eqnarray*}
where $\rho=\int d^3 v \,f$. As expected there is an infinite number of solutions for the parent bracket. A family of solutions is given by
\begin{equation}
\label{eq:MVparent}
\{F,G\}=\{F,G\}_{\rm VM}+\int d^3x \,\left( \nabla \cdot F_{\bf B} {\cal D} \nabla \cdot G_{\bf E}-\nabla \cdot F_{\bf E} {\cal D}^\dagger \nabla\cdot G_{\bf B}\right),
\end{equation}
where ${\cal D}$ is a linear operator independent of the field variables, so that the Jacobi identity is guaranteed by Morrison's lemma of Ref.~\cite{morr82}. This statement uses the fact that the Vlasov-Maxwell bracket has been made untainted; it would not be true if the original tainted Vlasov-Maxwell bracket~(\ref{eq:PBMV}) was considered instead of the Poisson bracket~(\ref{eq:PBMVproj}).

Now, if we apply the Dirac procedure on the extended Poisson bracket~(\ref{eq:MVparent}) with the primary constraint $\nabla \cdot\mathbf{E}-\rho$, we get the secondary constraint $\nabla \cdot\mathbf{B}$, and the reduced Dirac bracket is obtained from $\mathbb{J}_*=\mathcal{P}_*^\dagger\mathbb{J}\mathcal{P}_*$ where $\mathcal{P}_*$ is the Dirac projector~(\ref{eqn:defProjDir}). The Dirac projector can be explicitly computed. However,  in order to further simplify the computation of the Dirac bracket, we use the orthogonal projector since, as in the case of incompressible MHD (see Sec.~\ref{sec:MHDincomp}), it satisfies the same relation as the Dirac projector, i.e., $\mathbb{J}_*=\mathcal{P}_*^\dagger\mathbb{J}\mathcal{P}_*=\mathcal{P}_\bot^\dagger\mathbb{J}\mathcal{P}_\bot$, where
$$
	\mathcal{P}_\bot F_{\bm\chi}
	=(F_\mathbf{B}-\nabla \Delta^{-1}\nabla \cdot F_{\bf B}, F_\mathbf{E},  F_f) .
$$
This implies the expected result that the Vlasov-Maxwell bracket~(\ref{eq:PBMVproj}) is the Dirac bracket of the bracket~(\ref{eq:MVparent}) with Dirac constraints $(\nabla \cdot\mathbf{E}-\rho,\nabla \cdot\mathbf{B})$.

With the extended bracket~(\ref{eq:MVparent}), the Casimirs $(\nabla \cdot\mathbf{E}-\rho,\nabla \cdot\mathbf{B})$ of the Vlasov-Maxwell system now have  dynamics given by
\begin{eqnarray*}
	\frac{\partial }{\partial  t}\left(\nabla \cdot \mathbf{E} -\rho \right) 
	&=&
	\Delta\mathcal{D}^\dagger\nabla \cdot\mathbf{B}  ,
	\\
	\frac{\partial }{\partial  t}\nabla \cdot \mathbf{B} 
	&=&
	-\Delta\mathcal{D}\nabla \cdot\mathbf{E}  .
\end{eqnarray*}
These equations suggest two particularly interesting choices for our still undetermined operator $\mathcal{D}$. Defining $\mathcal{D}=\Delta^{-1}$ gives to $\nabla \cdot\mathbf{E}-\rho$ and $\nabla \cdot\mathbf{B}$ the dynamics of stationary waves when $\rho=0$, whereas defining $\mathcal{D}=(-\Delta)^{-1/2}$ gives them the dynamics of propagating waves. We note that these operators always act on divergences of vector fields. 

{\em Remark:} As a side note, we point out that the choice of ${\cal D}=(-\Delta)^{-1/2}$ naturally exhibits the operator $\nabla *:=\nabla (-\Delta)^{-1/2} \nabla \cdot$ which corresponds to $\nabla \times $ for the compressible part of a vector field. Indeed, the operator
$$
	-\nabla * \Delta^{-1}\nabla  * =\nabla \Delta^{-1}\nabla \cdot ,
$$
is the orthogonal projector onto the kernel of $\nabla \times $, just as $-\nabla \times \Delta^{-1}\nabla \times =1-\nabla \Delta^{-1}\nabla \cdot$ is the complementary projector onto the kernel of $\nabla  *$. With this choice, the resulting dynamical equations associated with the Poisson bracket~(\ref{eq:MVparent}) for the solenoidal and the compressible parts of the electromagnetic fields become independent and similar:
\begin{eqnarray*}
	& \square \mathbf{E}_S  = - \dot{\mathbf J}_S  , 
	\qquad
	& \square \mathbf{E}_C  = - \dot{\mathbf J}_C  , 
	\\
	& \square \mathbf{B}_S  =  \nabla \times \ {\mathbf J}_S  ,
	\qquad
	& \square \mathbf{B}_C  =  \nabla *\ {\mathbf J}_C  , 
\end{eqnarray*}
where $\square$ is the d'Alembert operator $\square=\partial ^2/\partial t^2 -\Delta$ and ${\bm \psi}_S$ is the solenoidal part of the vector field ${\bm\psi}$, i.e., ${\bm \psi}_S= - \nabla \times \Delta^{-1}\nabla \times  {\bm \psi}=(1-\nabla \Delta^{-1}\nabla) \cdot{\bm \psi}$ and ${\bm \psi}_C$ is its compressible part, which is ${\bm \psi}_C= - \nabla  * \Delta^{-1}\nabla  * {\bm \psi}= \nabla \Delta^{-1}\nabla \cdot{\bm \psi}$. In the absence of matter, the fields ${\bm \psi}_S$ and ${\bm \psi}_C$ propagate as independent free waves.

\subsection{From Vlasov-Maxwell to Vlasov-Poisson equations}

In order to obtain Vlasov-Poisson equations from the Vlasov-Maxwell equations we impose two constraints:
$$
{\cal Q}[f,{\bf E},{\bf B}]({\bf x})=({\bf B}-{\bf B}_0({\bf x}),\nabla \times {\bf E}),
$$
where ${\bf B}_0$ is a non-uniform background magnetic field. The operators $\hat{\cal Q}$ and $\hat{\cal Q}^\dagger$ are given by
$$
\hat{\cal Q}=\left( \begin{array}{ccc} 0 & \nabla \cdot & 0\\ 0 & 0 & 1
\end{array}\right), 
$$
and
$$
\hat{\cal Q}^\dagger=\left( \begin{array}{cc} 0 & 0\\ \nabla \cdot & 0\\ 0 & 1
\end{array}\right).
$$
The orthogonal projector ${\cal P}_\perp$ given by Eq.~(\ref{eqn:orth}) is given by
$$
{\cal P}_\perp=\left( \begin{array}{ccc} 1 & 0 & 0\\ 0 & \nabla \Delta^{-1} \nabla \cdot & 0\\
0 & 0 & 0 
\end{array}\right).
$$ 
Contrary to the orthogonal projector, the expression of the Dirac projector depends on the dynamics, and in particular on the Poisson matrix ${\mathbb J}$ which is given by
$$
{\mathbb J}=\left( \begin{array}{ccc} -[f,\cdot] & -\partial_{\bf v} f & 0\\
-f\partial_{\bf v} & 0 & \nabla \times \\
0 & - \nabla \times  & 0
\end{array}\right),
$$
where the small bracket $[\cdot,\cdot]$ is given by $[\cdot,\cdot]=[\cdot,\cdot]_c+[\cdot,\cdot]_{B_P}$ with these two brackets  given by Eqs.~(\ref{eqn:bc})-(\ref{eqn:bbp}).
The operator ${\cal A}$ is given by 
$$
{\cal A}=\left( \begin{array}{cc} 0 & (\nabla \times)^2\\ -(\nabla\times)^2 & 0
\end{array}\right). 
$$
The operator ${\cal A}$ is not invertible, but one can give an expression for ${\cal A}^{-1} \hat{\cal Q}$ by
$$
{\cal A}^{-1}\hat{\cal Q}=\left( \begin{array}{ccc} 0 & 0 & \Delta^{-1}(1-\nabla \Delta^{-1}\nabla\cdot)\\
0 & -\Delta^{-1}\nabla \times & 0
\end{array}\right).
$$
As a result, the Dirac projector is computed, 
$$
{\cal P}_*=\left( \begin{array}{ccc} 1 & 0 & 0\\
0 & \nabla \Delta^{-1} \nabla \cdot & 0\\
-\Delta^{-1}\nabla \times f\partial_{\bf v} & 0 & \nabla \Delta^{-1}\nabla \cdot 
\end{array}\right).
$$
We notice that both projectors ${\cal P}_\perp$ and ${\cal P}_*$ satisfy the 
equation ${\mathbb J}_*={\cal P}^\dagger {\mathbb J}{\cal P}$ and the Poisson matrix of the Vlasov-Poisson equations is given by 
$$
{\mathbb J}_*=\left( \begin{array}{ccc} 
-[f,\cdot] & -\nabla \Delta^{-1} \nabla \cdot \partial_{\bf v}f & 0\\
-\nabla \Delta^{-1} \nabla \cdot(f \partial_{\bf v}) & 0 & 0\\
0 & 0 & 0
\end{array}\right).
$$
It leads to the expression of the Poisson bracket, 
\begin{eqnarray*}
\{F,G\}_*= \int d^6z \,f[F_f-\Delta^{-1}\nabla\cdot F_{\bf E},G_f-\Delta^{-1}\nabla\cdot G_{\bf E}].
\end{eqnarray*}
Like for the incompressible MHD equations, even if the Dirac and orthogonal projectors are different, both of them can be used to compute the Dirac bracket from the Poisson matrix ${\mathbb J}$, the orthogonal projector being slightly simpler and more straightforward to compute. 

\section{Example 3: Quasi-neutrality as semi-local constraints}

In this section, we give an example of a set of physically relevant constraints where the orthogonal projector does not exist, and where the Dirac projector is a natural replacement for computing the constrained functional derivatives. We consider the following Vlasov equation with two species, ions and electrons, linearized about spatially homogeneous distribution functions $\alpha_i ({\bf v})$ and $\alpha_e ({\bf v})$. The equations for the phase space density {\em fluctuation} of ions $f_{\rm i}({\bf x},{\bf v})$ and electrons $f_{\rm e}({\bf x},{\bf v})$, are given by
$$
\dot{f_s}=-{\bf v}\cdot \nabla f_s +e_s \, \partial_{\bf v} \alpha_s \cdot\nabla \phi,
$$
where $\Delta \phi=-\sum_s e_s \int d^3v \,f_s$ and $e_s=\pm 1$ is the charge of the particles of each species $s={\rm i},{\rm e}$. 

The field variables are ${\bm\chi}({\bf z})= (f_{\rm i}({\bf z}), f_{\rm e}({\bf z}))$,  which are functions of ${\bf z}=({\bf x},{\bf v})$. The Poisson bracket, in this case, is defined by the Poisson matrix ${\mathbb J}$ given by
$$ 
{\mathbb J} = -\left(\begin{array}{cc}  [ \alpha_{\rm i}({\bf v}), \cdot\, ] & 0 \\0 & [ \alpha_{\rm e}({\bf v}), \cdot\, ]\end{array}\right),
$$
with $[ \alpha_s({\bf v}) , G ] =- \partial_{\bf v}\alpha_s \cdot \nabla G$.  According to Ref.~\cite{holm85} (see also Ref.~\cite{morr98}), the corresponding Hamiltonian can be found as the quadratic functional corresponding to the second derivative of the Hamiltonian (plus Casimirs) of the nonlinear Vlasov-Poisson system, evaluated at $f_s=\alpha_s ({\bf v})$. This is true for ion and electron densities close to equilibria that  are isotropic in velocity (like a Maxwellian for instance).  
We impose the set of two semi-local constraints 
$$
\bar{\cal Q}[{\bm\chi}]({\bf x}) = \left( \int d^3 v\, (f_{\rm i}-f_{\rm e}), \int d^3 v \, {\bf v} \cdot \nabla (f_{\rm i}-f_{\rm e})\right).
$$
The first component of the constraint is the quasi-neutrality. The second component is a secondary constraint associated with quasi-neutrality according to Dirac's theory of constrained Hamiltonian systems~\cite{Bhans76}. 
Since the constraints are linear with respect to the field variables, the operators $ \hat{\bar{\cal Q}}$ and $ \hat{\bar{\cal Q}}^\dagger $ are given by
$$  
{\hat{\bar{\cal Q}}}= \int {d^3 v} \left(\begin{array}{cc}  1 & -1 \\ {\bf v}\cdot \nabla &-{\bf v}\cdot \nabla \end{array}\right),~~~~\textrm{and } ~~~~~  \hat{\bar{\cal Q}}^\dagger  = \left(\begin{array}{cc}  1 & -{\bf v}\cdot \nabla \\ -1 & {\bf v}\cdot \nabla \end{array}\right).
$$
The operator $\hat{\bar{\cal Q}}^\dagger$ acts on functions of ${\bf x}$ only. 
In order to compute the Dirac projector, the matrix ${\cal A}=\hat{\bar{\cal Q}}{\mathbb J}\hat{\bar{\cal Q}}^\dagger$ needs to be computed
$$ 
{\cal A} = \left(\begin{array}{cc} 0 & ( \bar{\alpha}_{\rm i} + \bar{\alpha}_{\rm e})\Delta \\ -( \bar{\alpha}_{\rm i} + \bar{\alpha}_{\rm e})\Delta  & 2( {\bm \beta}_{\rm i} + {\bm \beta}_{\rm e})\cdot \nabla \Delta \end{array}\right),
$$
with $\bar{\alpha}_{s} =\int d^3 v\, \alpha_s$ and ${\bm \beta}_s =\int d^3 v\, {\bf v} \alpha_s$. 
This operator is invertible and its inverse is
$$ 
{\cal A}^{-1} =  \frac{1}{ \bar{\alpha}_{\rm i} + \bar{\alpha}_{\rm e}} \left(\begin{array}{cc} \displaystyle \frac{2( {\bm\beta}_{\rm i}+ {\bm\beta}_{\rm e}) }{ \bar{\alpha}_{\rm i} + \bar{\alpha}_{\rm e}}\cdot \nabla \Delta^{-1}  & -\Delta^{-1} \\  \Delta^{-1} & 0 \end{array}\right).
$$
Note that the operators ${\cal A}$ and ${\cal A}^{-1}$ act on functions of ${\bf x}$ and return a function of ${\bf x}$. 
The Dirac projector ${\cal P}_*$ has the form
$$
{\cal P}_*= 1 - \frac{\Delta^{-1} \nabla \cdot}{\bar{\alpha}_{\rm i} + \bar{\alpha}_{\rm e}} \int d^3 v'\, \,  \left({\bf v} +{\bf v}'-2\bar{\bf v}\right) \left(\begin{array}{cc}  [ \alpha_{\rm i} , \cdot ] & - [ \alpha_{\rm e} , \cdot ] \\ -[ \alpha_{\rm i} , \cdot ]   & [ \alpha_{\rm e} , \cdot ]  \end{array}\right),
$$
where ${\bar{\bf v}}=({\bm\beta}_{\rm i}+{\bm\beta}_{\rm e})/(\bar{\alpha}_{\rm i}+\bar{\alpha}_{\rm e})$.
Concerning the orthogonal projector, we note that $\hat{\bar{\cal Q}}  \hat{\bar{\cal Q}}^\dagger$ given by
$$
{\hat{\bar{\cal Q}}}  {\hat{\bar{\cal Q}}}^\dagger = 2\int d^3 v\, \left(\begin{array}{cc}  1 & -{\bf v}  \cdot\nabla \\ {\bf v} \cdot \nabla & - ({\bf v}\cdot \nabla)^2 \end{array}\right),
$$
does not exist since it is unbounded when it acts on functions of ${\bf x}$. As a consequence, the orthogonal projector cannot be a solution for the computation of constrained functional derivatives. Here a convenient choice is afforded by the Dirac projector.

\section*{Acknowledgments}
We acknowledge financial support from the Agence Nationale de la Recherche (ANR GYPSI). This work was also supported by the European Community under the contract of Association between EURATOM, CEA, and the French Research Federation for fusion study. The views and opinions expressed herein do not necessarily reflect those of the European Commission.  P J M was supported by U.S. Department of Energy contract DE-FG05-80ET-53088. The authors also acknowledge fruitful discussions with the \'Equipe de Dynamique Nonlin\'eaire of the Centre de Physique Th\'eorique of Marseille.


\section*{References}

\begin{thebibliography}{10}

\bibitem{zakh97} Zakharov V E and Kuznetsov E A 1997 {\it Phys. Usp.} {\bf 40} 1087

\bibitem{chan11} Chandre C, Morrison P J and Tassi E 2012 {\it Phys. Lett. A} {\bf 376} 737 

\bibitem{arno66} Arnold V I 1966 {\it Ann. Inst. Fourier} {\bf 16} 319

\bibitem{holm85} Holm D D, Marsden J E, Ratiu T S and Weinstein A 1985 {\it Phys. Rep.} {\bf 123} 1

\bibitem{holm98} Holm D D, Marsden J E and Ratiu T S 1998 {\it Adv. Math.} {\bf 137} 1

\bibitem{mars02} Marsden J E and Ratiu T S 2002 {\it Introduction to Mechanics and Symmetry} (Berlin: Springer-Verlag)

\bibitem{Dira50} Dirac P A M 1950 {\it Can. J. Math.} {\bf 2} 129

\bibitem{morr80a} Morrison P J and Greene J M 1980 {\it Phys. Rev. Lett.} {\bf 45} 790 

\bibitem{morr82e} Morrison P J and Greene J M 1982 {\it Phys. Rev. Lett.} {\bf 48} 569

\bibitem{morr82} Morrison P J 1982 \emph{Mathematical Methods in Hydrodynamics and 
Integrability in Related Dynamical Systems}, La Jolla Institute, 1981, edited by Tabor M and Treve Y M(1982 AIP Conference Proceedings {\bf 88} 13)

\bibitem{nguy99} Nguyen S and Turski L A 1999 {\it Physica A} {\bf 272} 48 

\bibitem{nguy01} Nguyen S. and Turski L A 2001 {\it Physica A} {\bf 290} 431

\bibitem{mars82} Marsden J E and Weinstein A 1982 {\it Physica D} {\bf 4} 394

\bibitem{morr80b} Morrison P J 1980 {\it Phys. Lett. A} {\bf 80} 383

\bibitem{bial84} Bialynicki-Birula I, Hubbard J C and Turski L A 1984 {\it Physica A} {\bf 128} 509

\bibitem{morr98} Morrison P J 1998 {\it Rev. Mod. Phys.}  {\bf 70} 467

\bibitem{Bhans76} Hanson A, Regge T and Teitelboim C 1976 {\em Constrained Hamiltonian Systems} (Roma: Accademia Nazionale dei Lincei)


\end{thebibliography}
\end{document}